\documentclass{mn2e}
\usepackage{graphicx}

\title[The recondite intricacies of Zeeman Doppler mapping]
      {The recondite\footnotemark[1] intricacies of Zeeman Doppler mapping}

\author[M.J.~Stift, F.~Leone and C.R.~Cowley]
       {M.J.~Stift$^1$, F.~Leone$^2$ and C.R.~Cowley$^3$\\
        $^1$Institut f{\"u}r Astronomie (IfA), Universit{\"a}t Wien,
            T{\"u}rkenschanzstrasse 17, A-1180 Wien, Austria\\
        $^2$Dipartimento di Fisica e Astronomia, Universit{\`a} die Catania,
            Sezione Astrofisica, Via S. Sofia 78, I-95123 Catania, Italy\\
        $^3$Department of Astronomy, University of Michigan, Ann Arbor,
            MI 48109-1042, USA}

\begin{document}

\date{Accepted 2011}

\pagerange{\pageref{firstpage}--\pageref{lastpage}} \pubyear{2011}

\maketitle

\label{firstpage}

 \begin{abstract}
   {We present a detailed analysis of the reliability of abundance and 
    magnetic maps of Ap stars obtained by Zeeman Doppler mapping (ZDM). 
    It is shown how they can be adversely affected by the assumption of a 
    mean stellar atmosphere instead of appropriate ``local'' atmospheres 
    corresponding to the actual abundances in a given region. The essence
    of the difficulties was already shown by Chandrasekhar's picket-fence 
    model. The results obtained with a suite of Stokes codes written in the 
    Ada programming language and based on modern line-blanketed atmospheres 
    are described in detail. We demonstrate that the high metallicity 
    values claimed to have been found in chemically inhomogeneous 
    (horizontally and vertically) Ap star atmospheres would lead to local 
    temperature structures, continuum and line intensities, and line shapes 
    that differ significantly from those predicted by a mean stellar atmosphere. 
    Unfortunately, past applications of ZDM have consistently overlooked 
    the intricate aspects of metallicity with their all-pervading effects. 
    The erroneous assumption of a mean atmosphere for a spotted star can 
    lead to phase-dependent errors of uncomfortably large proportions at 
    varying wavelengths both in the Stokes $I$ and $V$ profiles, making 
    precise mapping of abundances and magnetic field vectors largely
    impossible. The relation between core and wings of the H$\beta$ line 
    changes, too, with possible repercussions on the determination of 
    gravity and effective temperature. Finally, a ZDM analysis of the 
    synthetic Stokes spectra of a spotted star reveals the disturbing 
    differences between the respective abundance maps based on a mean 
    atmosphere on the one hand, and on appropriate ``local'' atmospheres 
    on the other. We then discuss what this all means for published ZDM
    results. Our discussion makes it clear that realistic local 
    atmospheres must be used, especially if credible small-scale 
    structures are to be obtained.
   } 
   \end{abstract}


\footnotetext[1]{{\bf recondite}: dealing with very profound, difficult, or 
    abstruse subject matter; requiring special knowledge to be understood 
    (http://dictionary.reference.com/browse/recondite)}

\section{Introduction}
\label{sec:intro}

Zeeman Doppler mapping (ZDM) has made considerable progress over recent years, 
thanks mainly to powerful new spectrographs which yield high S/N ratio polarised 
spectra in all 4 Stokes~$IQUV$ parameters. Whereas in the early days of Doppler 
mapping, intensity spectra only were used to derive abundance maps (see e.g. Rice 
et al. 1989), the addition of spectra in circular polarisation (Stokes~$V$) soon
opened up the possibility to simultaneously determine both the magnetic field 
structure and the abundance distributions of various chemical elements. However, 
right in the first days of ZDM Brown et al. (1991) gave a caveat: linear 
polarisation data (Stokes~$Q$ and $U$) would be needed to derive reliable 
magnetic maps for Ap stars. Still, even into the new millennium, Doppler maps 
of magnetic stars have been published, based on Stokes~$I$ and $V$ only 
(Kochukhov et al. 2002, L{\"u}ftinger et al. 2010), although instruments capable 
of determining accurate $Q$ and $U$ profiles have by now been developed. 
Numerous Doppler maps of magnetic stars even rely solely on Stokes~$I$ (Piskunov 
et al. 1998, Kuschnig et al. 1998, L{\"u}ftinger et al. 1998, 2003). The reader 
of the articles cited will surely note that elemental over-abundances resulting 
from Doppler imaging frequently appear quite unrealistic. Consider for example 
the results for $\iota$\,Cas where according to Kuschnig et al. (1998) in parts 
of the atmosphere $\log{({N_{Cr}}/{N_{tot}})} = -1.50$. Similarly, for 
$\kappa$\,Psc, Piskunov et al. (1998) find that over a considerable part of the 
surface $\log{({N_{Fe}}/{N_{tot}})} = -1.81$, while Cr, according to their 
figure~1, varies between $-6.09$ and $0.27$! For comparison, the solar
abundance is $-5.6$ (Asplund et al. 2009).The question not only arises how 
possibly a star can contain more chromium than hydrogen in parts of its 
atmosphere but also how such an enormous over-abundance could build up and 
whether such an atmosphere could ever remain stable. Note that the modelling 
of equilibrium stratifications in Ap stars (LeBlanc et al. 2009, Alecian \& 
Stift 2007) has not so far yielded a single case with such extreme iron or 
chromium abundances.  Could these excessive values not rather be due to serious 
shortcomings in the ZDM procedure?

Fortunately, high quality Stokes~$QU$ profiles have now become available and this 
might be thought to ensure at last uniqueness of the magnetic and abundance maps.
Based on recently obtained Stokes~$IQUV$ profiles, a major revision both of the 
previously published magnetic geometry and of the elemental abundance distributions 
of the famous Ap star $\alpha^2$\,CVn has taken place. The new maps appear to 
validate the analysis of Brown et al. (1991). Whereas the Stokes~$I$ and $V$ based 
analysis by Kochukhov et al. (2002) (=K02) yielded a minimum field of $-6.5$\,kG 
and a maximum field of $+5.1$\,kG, these values have now become $-3.5$\,kG and 
$+3.5$\,kG respectively (Kochukhov \& Wade 2010) (=K10); we see also substantial 
changes in the structure both of the meridional and the azimuthal field. The 
respective field distributions from K02 and K10 at phases 0.20 and 0.40 are 
strikingly different and it is quite surprising to see that a sharp maximum 
in field strength is accompanied by an almost horizontal inclination of the 
magnetic field. 

The K10 iron maps bear little resemblance to the K02 maps, and there are a 
few noticeable differences in the chromium maps; there is no obvious correlation 
between field direction or strength and the abundance features. 
A 2.3\,dex amplitude in the Fe abundances has increased to a
staggering 4.9\,dex, a fact that cannot be explained by magnetic intensification 
(see Stift \& Leone 2003). However, there are not only these differences between 
the old and the new maps that are distressing, but there is also the question of
(optimum) regularisation. In an ill-posed problem such as ZDM, regularisation is
necessary to obtain meaningful and hopefully unique maps of elemental abundances 
and of the magnetic field vector; these maps should combine minimum structure
(complexity) with a good fit to the observed Stokes~$IQUV$ profiles. In the past, 
maximum entropy regularisation has been used extensively (see e.g. Vogt et al. 
1987), but nowadays it is largely replaced by Tikhonov regularisation which 
minimises the sum of the squared differences between the unknowns (abundances, 
magnetic field components) of any combination of 2 surface elements (see eq.\,5 
of K10). Please keep in mind that both Tikhonov and maximum entropy regularisation 
are purely mathematical constraints which do not necessarily reflect the physical 
reality: radiatively driven diffusion in magnetic stellar atmospheres might not 
lead to smooth but rather to patchy surface structure (Alecian \& Stift 2010).

Even with all 4~Stokes parameters used by K10, the magnetic geometry depends in 
a sensitive way and to an uncanny degree on the value of the regularisation 
parameter. Obviously, the smaller the latter gets, the more fine structure in the 
magnetic field appears, but in addition, changes in the regularisation parameter 
make field maxima not only change appreciably in value but these maxima also appear 
to travel over substantial distances to new locations (compare figures 6 and 8 in 
K10). In that context the reader of the K10 paper, having access to abundance maps 
for only 2 values of the regularisation parameter, is certainly entitled to ask 
himself whether this travel proceeds smoothly or rather in erratic jumps. The reader 
will also note that Stokes $Q$ and $U$ data for a mere 3 or 4 out of 20 phases 
are mainly responsible for the complete revision of the surface abundance and field 
strength maps (this will be discussed in section~7) while there are still a lot of 
profiles that have not been fitted to within the estimated observational errors.

\begin{figure}
\includegraphics[width=84mm]{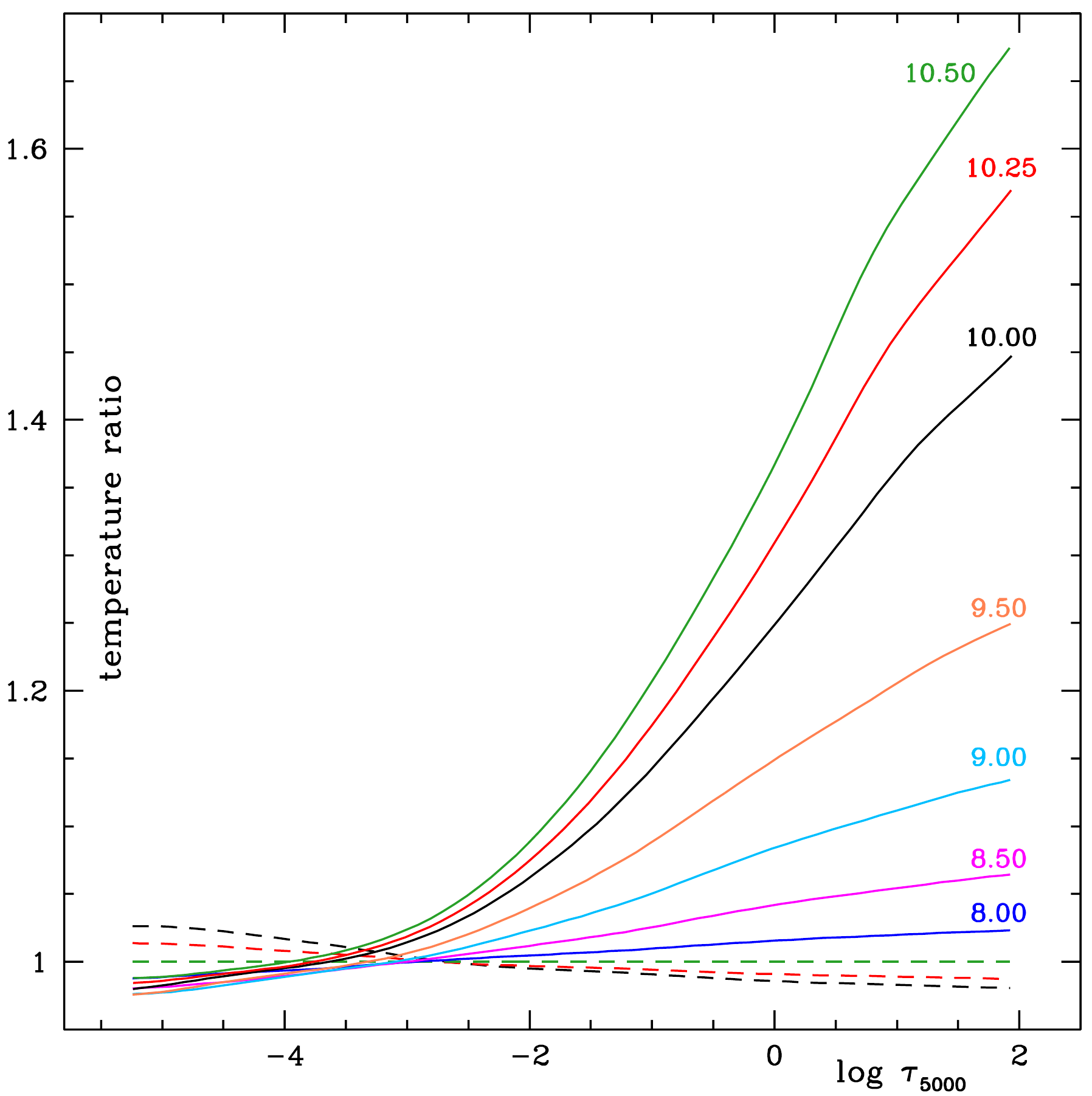}
\caption{Variations, as a function of iron abundance, in the run of 
temperature with $\log \tau_{5000}$ in an Atlas12 model atmosphere with 
$T_{\rm eff} = 12000$\,K, $\log g = 4.0$ (all other elements have solar 
abundances). The curves pertain to iron abundances ranging from 
$A({\rm Fe}) = 6.50$ to $10.50$ and are plotted relative to the temperature 
profile of the $A({\rm Fe}) = 7.50$ model.}
\label{atmos}
\end{figure}

Apparently the method used by K10 is not robust, since the results depend 
too strongly on the value of the regularisation parameter. As K10 have shown 
themselves in their figures 6 and 8, the value of the regularisation parameter 
not only determines the resolution of the maps -- as one would ideally expect --
but to an uncomfortable degree also their large-scale structure. Is this fact
due to some numerical instability and/or to weird behaviour of the Tikhonov 
regularisation, or is it rather indicative of something even more disturbing, 
i.e. do we have to admit that the input physics used so far by K10 are seriously 
inadequate? Could it be that the ZDM code yields a beautiful, but incorrect and 
unphysical solution that also depends on erroneous assumptions as to, for example, 
the atmospheric structure? This is what we will explore in the following sections.

\section{The tools}

Any thorough investigation of ZDM and its input physics requires a set of powerful
numerical codes. One has to be able to calculate highly accurate Stokes profiles
for stellar atmospheres permeated by strong magnetic fields of various geometries.
There is the need for opacity sampled stellar atmospheres which allow for 
arbitrary, perhaps even vertically stratified, elemental abundances. The spectra of
spotted stars have to be calculated using at least 2 different atmospheric models
(for a single spot) but more flexible and realistic abundance patterns may have to
make use of many more model atmospheres, up to 10 and even more. Finally it is 
useful to be able to verify how (Zeeman) Doppler maps reflect or fail to reflect 
any spotted structure adopted in the synthesis of Stokes profiles of magnetic 
stars.

All but one of the codes used in our investigation are based on C{\sc ossam} which
stands for ``{\bf Co}dice per la {\bf s}intesi {\bf s}pettrale nelle {\bf a}tmosfere
{\bf m}agnetiche'' described herewith. Parts of C{\sc ossam} can be traced back to 
A{\sc drs3} (Chmielewski 1979) which itself is an evolved version of the Fortran 
adaptation by Peytremann et al. (1967) of the Algol\,60 code A{\sc nalyse\,65} by 
Baschek et al. (1966). C{\sc ossam} and all the other codes are written in Ada, an 
object-oriented language whose syntax derives from the Algol family of languages and 
which is used mainly for safety-critical applications such as flight software but is 
also eminently suitable for scientific computing. Ada makes it possible to achieve 
a remarkably high degree of portability, its encapsulated software modules 
(``packages'') allow extensive verbatim software reuse, and there are unique 
language constructs (``tasks'') for concurrent processing which can be employed 
for highly efficient parallel computing with very little synchronisation overhead 
thanks to ``protected types''. Finally it should be mentioned that free Ada
compilers of the same quality as those provided for EADS or Boeing are provided
free of charge and are included in every Linux distribution. For more details
see Stift (1998, 2000) and Wade et al. (2001).

C{\sc ossam} is a code that calculates full Stokes~$IQUV$ profiles in
local thermodynamic equilibrium (LTE) either 
for the solar case (at a given point on the solar surface) or for the stellar case 
(integral over the visible hemisphere). Atomic transition data are usually taken 
from the VALD database (Piskunov et al. 1995, Kupka et al. 1999), and often include 
the constants for radiation damping, Stark broadening and van der Waals broadening.
When these constants are unavailable, classical radiation damping and Uns{\"o}ld
van der Waals broadening are taken, together with Stark broadening according
to Gonzalez et al. (1995). For the Voigt and Faraday profiles of the metal lines
the complex rational function given by Hui et al. (1978) is used; the hydrogen 
line profiles are calculated by means of the approximation found in T{\sc lusty}
(Hubeny \& Lanz 1995). Zeeman splittings are calculated from the J-values of 
the lower and upper energy levels respectively and their Land{\'e} factors; in
case these data are lacking, a simple Zeeman triplet is assumed. In order to
establish the line absorption matrix in the polarised radiative transfer equation, 
full opacity sampling is carried out, separately for the $\sigma_{-}, \sigma_{+}$ 
and $\pi$ components. The required continuous opacities come from the A{\sc tlas12} 
(Kurucz 2005) routines translated to Ada and encapsulated in a reusable 
multi-purpose package. Special care is applied to the treatment of the Lyman,
Balmer and Paschen discontinuities according to the level dissolution probability
method of Hubeny et al. (1994). The formal solution to the polarised RTE is
carried out using the Zeeman Feautrier method (Auer et al. 1977, Alecian \&
Stift 2004) or the DELO method (Rees et al. 1989). Spatial integration of the
local Stokes profiles over the visible hemisphere can take advantage of an
optimum grid due to Stift (1985). Opacity sampling, the most time-consuming part 
of the code, is entirely and efficiently parallelised and so allows one to take 
full advantage of the power of modern symmetric multi-processing (SMP) machines 
of  1-48 cores and more, all of this without having recourse to MPI or to 
related software libraries.

C{\sc ossam}S{\sc pot} is a code derived from C{\sc ossam} that calculates 
full Stokes~$IQUV$ profiles for stars doted with as many as 10 spots. As we 
shall show below, atomic transition data have to be accommodated not just 
for 1 atmosphere but for up to 10 atmospheres (or even more). Thanks to the 
object-orientation of Ada this can be done in a straightforward way through 
the use of {\em generic packages} which constitute {\em templates} (in the 
C++ diction) that can be {\em instantiated} with actual parameters. It thus 
becomes possible to model e.g. the spectrum of a star with a number of spots 
each of which is made up of concentric rings of different abundance and/or 
temperature. Clearly, spotted stars require a grid that is different from the 
observer-centred grid used in C{\sc ossam}: here we have a co-rotating grid 
where the whole surface of the star is split into elements of about equal area 
(for a typical such grid see e.g. figure~1 in Vogt et al. 1987).

For the stellar atmosphere models we rely on the A{\sc tlas12} code (Kurucz 2005),
translated to Ada by Bischof (2005) and offering a restricted number of 
Kurucz's original options; the code has subsequently been thoroughly debugged 
by Stift and now offers most of the original options. A{\sc tlas}A{\sc da} makes 
provisions for stratified atmospheres and thanks to the parallelisation of 
the opacity sampling part, can provide excellent frequency resolution. Apart 
from these 2 improvements and from better defined interfaces to the subprograms, 
the input physics are identical to those of the official Fortran version.

Finally, the Doppler mapping code by Stift (1996) has been completely rewritten
and is now based on the latest version of C{\sc ossam}. At the time of writing, 
C{\sc ossam}D{\sc oppler} derives abundance maps from Stokes~$I$ profiles only; 
the magnetic field geometry is taken into account but has -- for our tests -- 
to be the same as the input geometry to C{\sc ossam}S{\sc pot}. Maximum entropy 
regularisation is combined with a simple gradient search, ensuring convergence 
to the desired high-quality fit to the intensity profile within 6-10 iterations 
in favourable circumstances.

\section{The picket-fence model and Atlas12 stellar atmospheres}

The repercussions of line blanketing on the structure of a stellar atmosphere
were treated in an analytical way  by Chandrasekhar (1935). His so-called
picket-fence model is based on the assumptions of a frequency-independent
continuum opacity, of lines distributed randomly and uniformly over the
spectrum, and of a constant opacity ratio between the lines (with square 
profiles) and the continuum -- for a very readable presentation of this method 
see Mihalas (1970). Chandrasekhar showed that with increasing line opacity, 
the temperature gradient increases, and the temperature at the upper boundary
of the atmosphere decreases. Later numerical approaches allowed for more
realistic spectral line distributions, starting with 30000 lines for Procyon
(Strom \& Kurucz 1966) and going up to millions of lines (Kurucz \& Bell 1995); 
they give a very detailed idea of how important line blocking and backwarming 
can become and how these affect the structure of a stellar atmosphere.

\begin{figure}
\includegraphics[width=84mm]{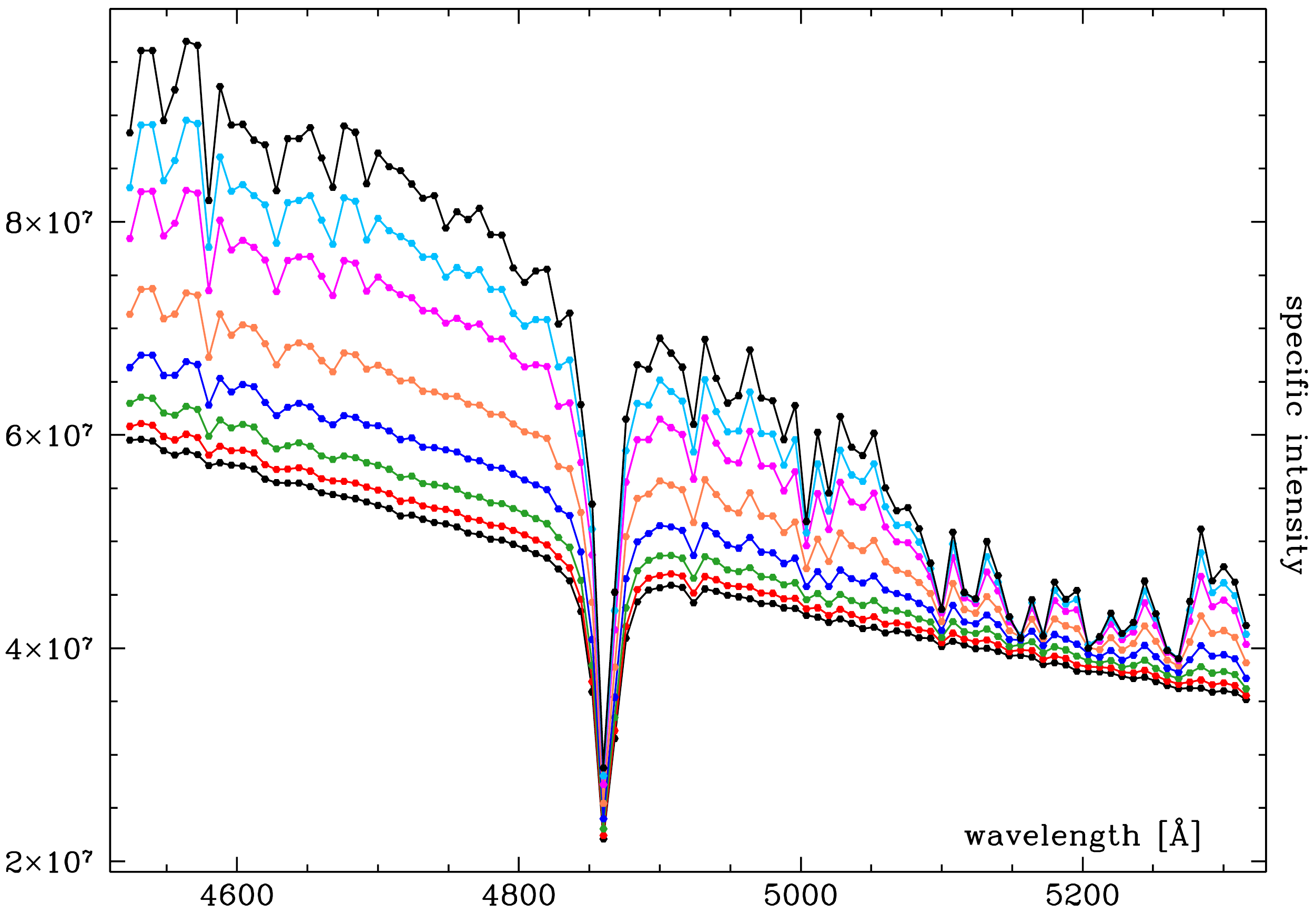}
\caption{Line blanketed spectra for Atlas12 model atmospheres with 
$T_{\rm eff} = 12000$\,K,  $\log g = 4.0$ and varying iron abundances. For all 
other elements, solar abundances have been assumed. Means of the specific 
intensity at the centre of the disk (in ergs\,cm$^{-2}$\,s$^{-1}$\,Hz$^{-1}$\,sr$^{-1}$)
have been calculated over 8\,{\AA} intervals. The curves pertain -- from bottom 
to top -- to $A({\rm Fe}) = 7.50, 8.00, 8.50, 9.00, 9.50, 10.00, 10.25, 10.50$.}
\label{blanket}
\end{figure}

Chandrasekhar's picket-fence model is an astrophysical classic but nobody seems 
to have spent much thought on it in the context of ZDM. So let us investigate 
what fully line-blanketed atmospheres calculated with A{\sc tlas12} (Kurucz 2005) 
look like for the extreme abundances, especially of iron, claimed by K10 for 
$\alpha^2$\,CVn. For this purpose we have established a grid of models with 
$T_{\rm eff} = 12000$\,K, $\log g = 4.0$, and with iron abundances that cover a wide 
range, viz. $A({\rm Fe}) = 6.50, 7.00, 7.50, 8.00, 8.50, 9.00, 9.50, 10.00, 10.25$
and $10.50$ (on a scale where the hydrogen abundance is 12.00). For simplicity's sake 
we took solar abundances for all the other elements, allowing us to single out the 
effects of just 1 parameter. Convergence of the models with large over-abundances 
turned out to be excruciatingly slow at times and it also proved necessary to 
proceed in small steps from $A({\rm Fe}) = 9.00$ to $A({\rm Fe}) = 10.00$ and 
beyond in order to obtain any convergence at all. Thanks to the parallelised Ada 
version of A{\sc tlas12} we could afford satisfactory sampling with 120000 frequency 
points. Fig.\,\ref{atmos} clearly illustrates how the temperature gradient 
steepens with increasing metallicity, and how at $\log \tau_{5000} = 2.0$, the 
bottom of the atmosphere gets hotter whereas the top at $\log \tau_{5000} = -5.3$ 
becomes cooler, at least up to moderate metallicities. Let us point out that the 
differences between the curves for $A({\rm Fe}) = 6.50$ and $A({\rm Fe}) = 8.00$ 
are smaller than the differences between the curves for $A({\rm Fe}) = 10.00$ and 
$A({\rm Fe}) = 10.25$! In order to illustrate the vast changes in line blocking 
between $A({\rm Fe}) = 7.50$ and $A({\rm Fe}) = 10.50$ over the interval covered 
by the metal lines used in K10 we plot in Fig.\,\ref{blanket} Stokes\,$I$ specific
intensity means over 8\,{\AA} intervals. At these wavelengths, the continuum of 
the $A({\rm Fe}) = 10.50$ atmosphere lies considerably above the continuum of the
$A({\rm Fe}) = 7.50$ atmosphere in order to ensure constant total flux for the
given effective temperature in the presence of heavy blanketing. Note the nice 
depression around 5200\,{\AA} which develops for $A({\rm Fe}) > 10.00$ and which 
is probably related to the well known feature used to define a peculiarity index 
(Maitzen 1976). Given the large differences in atmospheric structure, can we 
expect spectral line shapes to be insensitive to metallicity effects?

\section{Line shapes and simple arithmetics}

Notwithstanding the ground-breaking work of Chandrasekhar (1935), differential 
line-blanketing -- between spot and photosphere -- in stellar atmospheres has never 
been taken into account by Doppler mapping experts. Especially for recent work this
seems a bit strange since Khan \& Shulyak (2007) have shown the non-negligible
effects of Fe and Cr over-abundances on stellar atmospheres and on abundance
estimates. True, for very small abundance differences over the stellar surface it
is absolutely tolerable to approximate the ``local atmospheres'' by an atmosphere 
corresponding to the mean metallicity (but see below for the definition of 
``very small''). However, in the light of the findings of Khan \& Shulyak (2007) 
it is certainly time to challenge this mean atmosphere assumption in those cases 
when iron abundances attain and exceed $A({\rm Fe}) = 9.50$, and to show that line 
profiles calculated with the corresponding high-metallicity atmospheres differ 
significantly from those calculated for example with a $A({\rm Fe}) = 7.50$ or even 
a $A({\rm Fe}) = 8.50$ atmosphere.

\begin{figure}
\includegraphics[width=84mm]{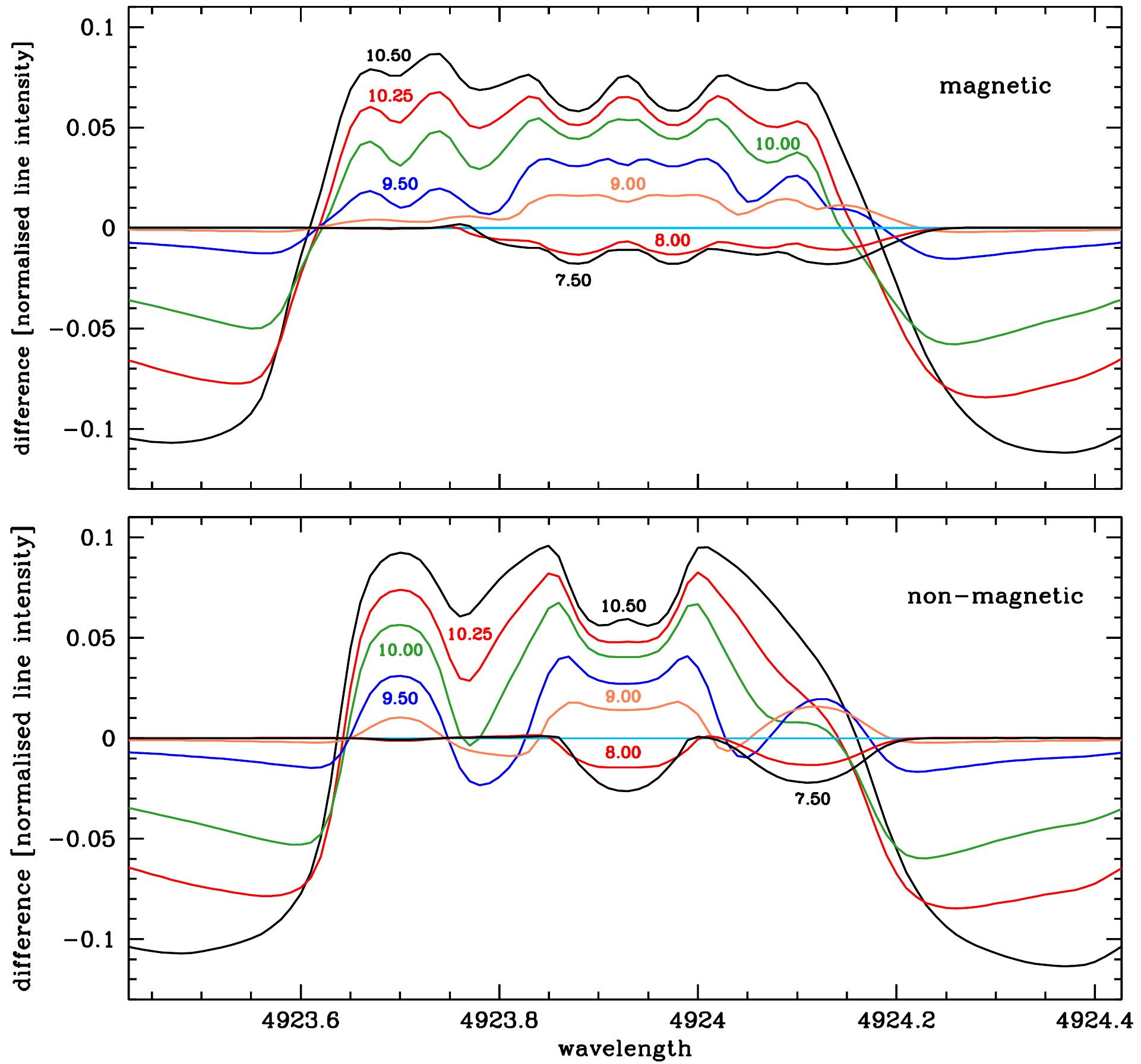}
\caption{Abundance-dependent, differential shapes of the normalised Stokes~$I$ 
profile of the Fe\,{\sc ii} line at 4923.927\,{\AA}. The plots give, for a number
of iron abundances $A({\rm Fe})$, the differences $I_{850}^{A} - I_{A}^{A}$ at the 
centre of the stellar disk; the subscript denotes the iron abundance assumed in 
the establishment of the stellar atmosphere later used for the spectral synthesis, 
the superscript the actual iron abundance adopted. Both the non-magnetic case 
(lower panel) and the magnetic case (upper panel) with field strength 4400\,G and 
$90\degr$ angle between field vector and line-of-sight are shown. The model 
atmospheres are the same as presented in Fig.\,\ref{blanket}.}
\label{shape}
\end{figure}

For this purpose, Fig.\,\ref{shape} displays, for the $\lambda$\,4924 iron line 
(plus actual blending lines) at the centre of the stellar disk, normalised 
intensity differences $I_{850}^{A} - I_{A}^{A}$ for various adopted Fe abundances 
$A({\rm Fe})$. The subscript stands for the iron abundance adopted in the 
establishment of the stellar atmosphere, the superscript for the actual iron 
abundance assumed for the spectral synthesis. The abundance grid for this test 
covered $A({\rm Fe}) = 7.50, 8.00, 8.50, 9.00, 9.50, 10.00, 10.25, 10.50$. In 
the lower panel we plot the non-magnetic intensity differences, in the upper 
panel the differences for a 4400\,G field inclined by $90\degr$ with respect to 
the line-of-sight. For $A({\rm Fe}) = 10.00$ the differences attain about 5\% 
of the continuum both in the non-magnetic and in the magnetic case, and they 
get substantially larger (more than 10\%) with increasing iron abundance. In 
view of these results it is difficult to understand how K10 (in section\,3) 
can come to the conclusion that {\em ``the local line profiles are sensitive 
to model structure effects to a much smaller degree than to changes of abundance 
or magnetic field''}. It also emerges from these plots that for ultra-high 
abundances, it becomes more or less impossible to correctly model the line core. 
The Atlas12 atmospheres -- for good reasons -- do not go further outward than 
about $\log \tau_{5000} = 5.0\,10^{-6}$ and this turns out to be no longer 
sufficient for extremely strong lines. While this problem hardly shows up when 
using a $A({\rm Fe}) = 8.50$ atmosphere, it becomes an object of preoccupation 
with the correct atmospheric models.

The whole problem with the adoption of a single atmospheric model for the mapping 
of spotted stars with large differences in abundances over the surface boils down 
to the inequality \\
$(S_1 I_{A1} / C_{A1} + S_2 I_{A2} / C_{A1}) \neq $\\
\indent\indent\indent\indent $(S_1 I_{A1} + S_2 I_{A2}) / (S_1 C_{A1} + S_2 C_{A2})$\\
where $I$ stands for the line intensity, $C$ for the continuum intensity. $S_1$ 
and $S_2$ are the fractions of the visible surface taken by spot and remaining
atmosphere respectively, $A1$ and $A2$ are the respective abundances inside and 
outside the spot. The left-hand side corresponds to the approximation used in
all previous ZDM work, whereas the right-hand side gives the correct formula for
the observed normalised intensity. It comes as no surprise that the consequences 
of not taking into account the metallicity-dependent line shapes and absolute 
line intensities can be  quite serious as we shall demonstrate further in the 
following sections.

\begin{figure}
\includegraphics[width=84mm]{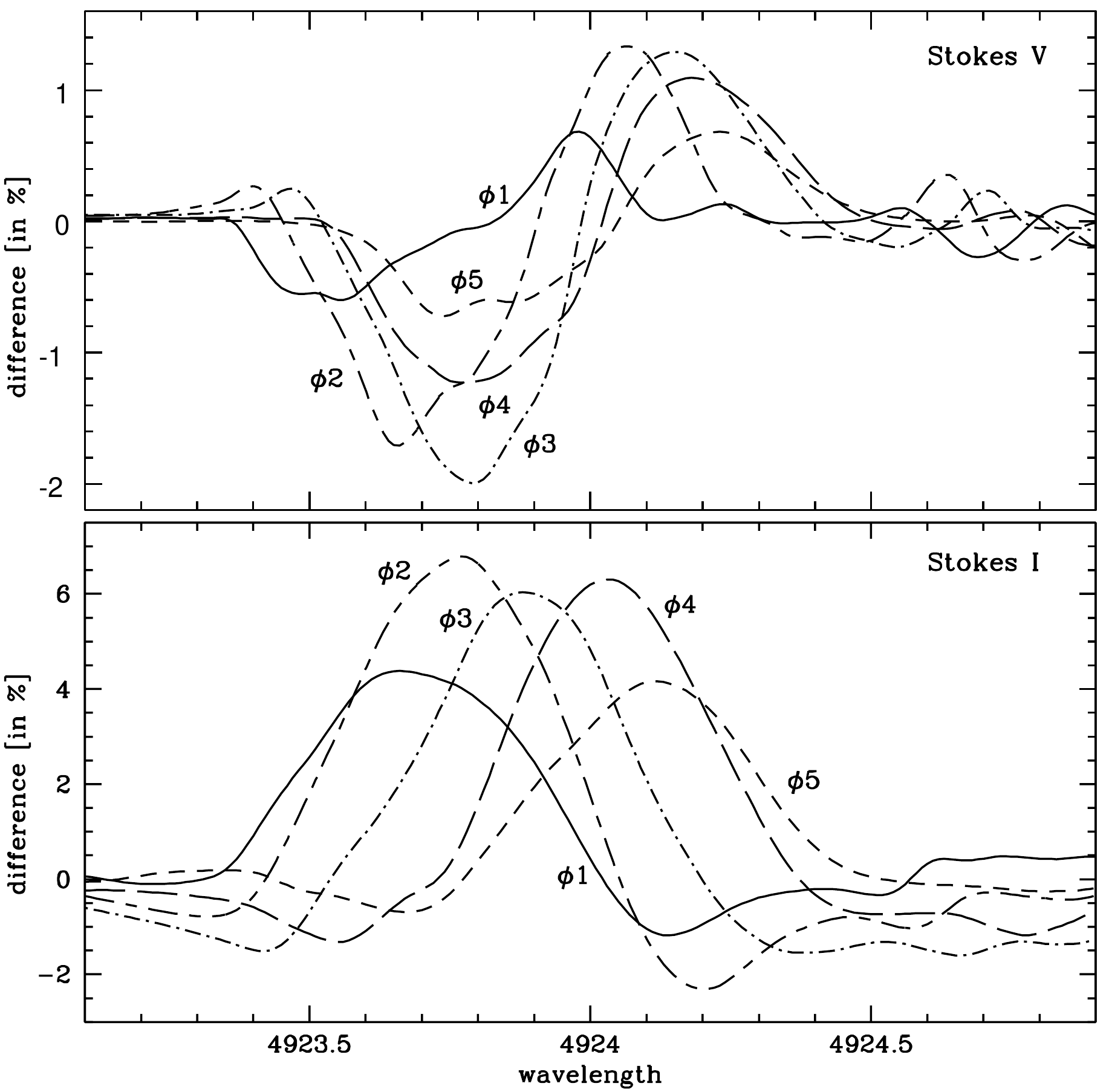}
\caption{Differences between the Stokes~$I$ and $V$ profiles of the Fe\,{\sc ii} 
line at 4923.927\,{\AA} of a rotating star with a single spot, calculated with 
the correct (2) model atmospheres for the star and its spot on the one hand, 
with a mean model for star and spot on the other hand. The plot shows the 
normalised differences for 5 phases ($\phi 1 = 0.000$, $\phi 2 = 0.125$, 
$\phi 3 = 0.250$, $\phi 4 = 0.375$, $\phi 5 = 0.500$) in the sense mean minus 
correct. A spot of $50\degr$ radius is situated at a latitude of $40\degr$ 
and displays a 1.5\,dex over-abundance of iron with respect to the 
$A({\rm Fe}) = 8.50$ abundance over the rest of the star. The inclination of 
the star is $60\degr$, the respective magnetic field extrema are 1500\,G and 
5500\,G. At $\phi 1$ and $\phi 5$ the spot is near the limb, at $\phi 3$ it 
is almost centred.}
\label{Stokes-IV}
\end{figure}

\section{Abundances and magnetic field strengths}

What can we learn from the plots given in the section above as far as the 
determination of abundances is concerned? Extreme Fe abundances in localised
spots as claimed by K10 for $\alpha^2$\,CVn -- there would be 3 iron atoms 
for every 100 hydrogen atoms -- seem a bit unrealistic but are no obstacle 
to modelling with A{\sc tlas12} and with C{\sc ossam}. We have calculated the 
full Stokes parameters of a spotted star with $v \sin i = 18$\,km\,s$^{-1}$, 
inclination $i = 60\degr$, and covered by a large spot of $50\degr$ radius 
at $40\degr$ latitude. For the spot we adopted $A({\rm Fe}) = 10.00$, for 
the remaining part of the star $A({\rm Fe}) = 8.50$. One set of spectra was 
calculated with a $A({\rm Fe}) = 8.50$ atmosphere throughout, the other set 
is based on the actual value, $A({\rm Fe}) = 10.00$, inside the spot.

The lower panel of Fig.\,\ref{Stokes-IV} displays -- for 5 different rotational 
phases -- the differences between the normalised Stokes~$I$ spectrum obtained 
with the same atmosphere all over the star (as done by K02/K10 and henceforth 
denoted as the ``usual'' spectrum) and the correct spectrum. These differences 
with a maximum of about 7\% -- which can by no means be considered small or 
negligible in view of a central line intensity of about 40\% of the continuum 
-- change their positions with phase due to the rotation. In a ZDM calculation, 
based on the ``usual'' global atmospheric model, this would translate to 
appropriately placed spurious over-abundances, a problem which we shall discuss 
later in more detail.

The upper panel of Fig.\,\ref{Stokes-IV} reveals that polarisation is affected 
too. The tilted and eccentric dipole model adopted leads to respective minimum 
and maximum field strengths of 1500 and 5500\,Gauss. The amount by which Stokes~$V$ 
can be in error reaches some 2\% which has to be put in relation to the maximum 
signal of about 10\%. Again, the differences change position with phase and 
simulate spurious magnetic field structure.

\begin{figure}
\includegraphics[width=84mm]{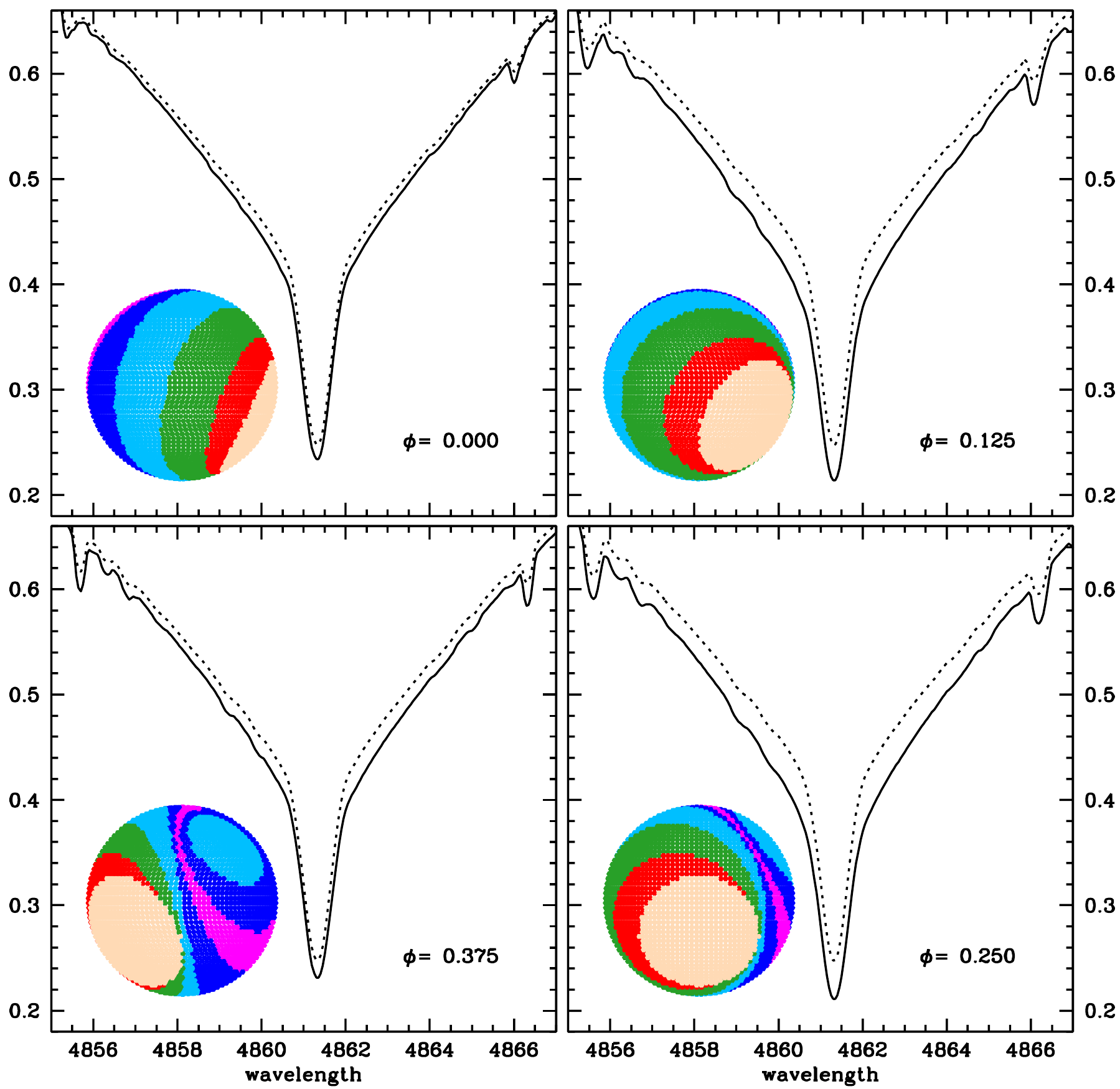}
\caption{Surface abundance distributions and corresponding H$\beta$ profiles 
at 4 different rotational phases of a star with a complex spot structure and 6 
different iron abundances $A({\rm Fe})$, viz. 7.50 (magenta), 8.00 (blue), 8.50 
(sky-blue), 9.00 (green), 9.50 (red), and 10.00 (peach). The star is seen 
equator-on. The full lines pertain to the normalised Stokes~$I$ profiles 
calculated at the phases indicated with the correct atmospheres corresponding 
to the adopted abundances, the dotted lines are based on the ``usual'' 
$A({\rm Fe}) = 8.50$ atmosphere and remain constant.}
\label{beta}
\end{figure}

\section{The H$\beta$ line}

The profile of the H$\beta$ line, together with the profiles of other Balmer lines, 
are used for the determination of the effective temperature and the surface gravity 
of a star. If, as shown above, the shapes of iron lines change with the metallicity 
of the stellar atmosphere, what will happen to the Balmer lines in a spotted star and 
to the stellar parameters derived from them? We have tried to figure this out at least 
partially by calculating phase-dependent spectra for a star with 8 abundance regions 
representing 6 different iron abundances. The results are shown for 4 phases in 
Fig.\,\ref{beta}, together with the corresponding abundance distributions over the 
visible stellar hemisphere. The dotted curves correspond to the H$\beta$ line derived 
with the ``usual'' $A({\rm Fe}) = 8.50$ atmosphere -- it must remain constant 
irrespective of phase since we do not consider Zeeman splitting of hydrogen lines -- 
the full lines are based on the 6 correct model atmospheres.

It is immediately clear that the H$\beta$ line does not remain unaffected by 
high-metallicity spots, in accord with the findings of Leone \& Manfr{\'e} (1997). 
In the particular case presented here, the core is deeper by up to 5\% of 
the continuum and the inner wings larger, something that may possibly have 
repercussions on the determination of stellar parameters. The outer wings turn 
out to change by less than 1\% at wavelengths more than 10\,{\AA} distant from 
the line centre. It remains to be seen if these profile changes are related to 
the core-wing anomaly discussed by Cowley et al. (2001).

\section{The detectability of magnetic fine structure}

Kochukhov \& Wade (2010) claim to have found ``high-contrast surface distributions" 
of Fe and Cr together with ``small-scale magnetic structures" and a global field 
dominated by ``high-contrast magnetic spots". Looking at their figure~6 we indeed see 
``the high-contrast structure of the field strength at smaller spatial scales'' 
consisting in particular of 2 very intense magnetic spots of surprisingly small 
extension; at the same time their figure~9 does not reveal any similar small-scale 
structure in the abundance distributions of Fe or Cr. The magnetic spots exhibit 
diameters which lie in the approximate range of $15-25\degr$, lines of constant 
field strength are separated by as little as $8\degr$ (for a difference of 0.5\,kG), 
and lines of constant abundances by about $6\degr$ (for a difference of 0.5\,dex in 
abundances). 

\begin{figure}
\includegraphics[width=84mm]{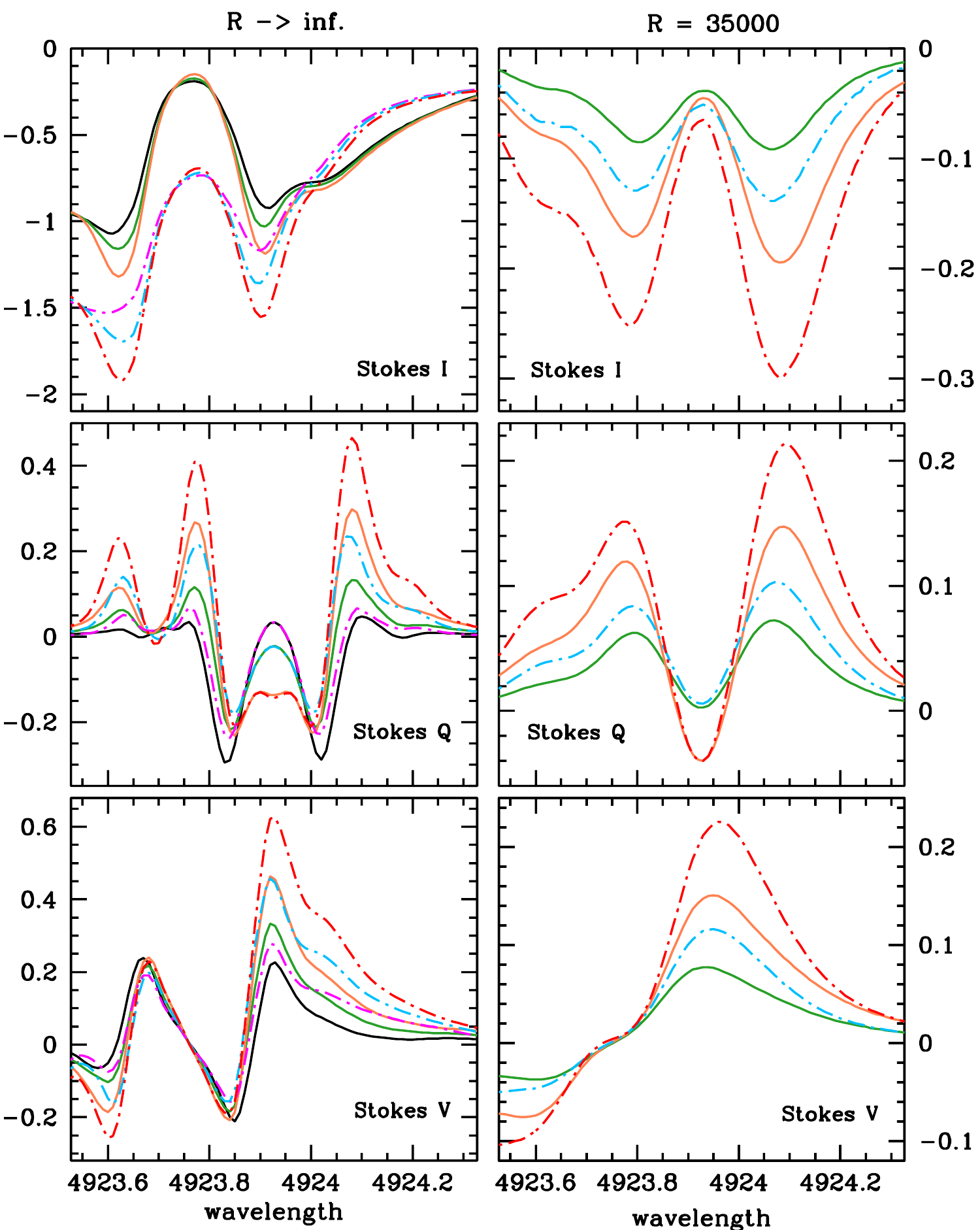}
\caption{Detectability of magnetic fine structure in a star with a spot 
of $20\degr$ diameter and $A({\rm Fe}) = 10.00$ abundance. Outside the spot 
an abundance of $A({\rm Fe}) = 8.50$ is assumed. The full curves pertain to 
the ``usual'' model, the dash-dotted lines to the correct model. The panels 
to the left display Stokes profiles relative to those of a spotless star,
calculated for essentially infinite spectral resolution, and with field 
strengths inside the spot multiplied by factors of 1.00 (black, magenta), 
1.50 (green, sky-blue) and 2.00 (peach, red) respectively. Units are \% of 
the continuum or \% polarisation. The panels to the right display profiles 
relative to those of a star with no field enhancement in the spot, convolved 
to the resolution of the K10 spectra. Note the difference in scale!
}
\label{fine}
\end{figure}

Before embarking on the numerical modelling of the signal of a high-contrast 
magnetic spot, let us shortly dwell on just 2 appropriate observational 
considerations. Looking at figure~7 in K10, it appears that a feature as small 
as $20\degr$ is resolved which in terms of spherical harmonics would correspond 
to $m = 18$. However, at the same time figure~12 in K10 goes no further than 
$l = m = 10$ with hardly any power left in this moderately high harmonic. This 
raises first doubts as to the actual detectability of small-scale structures, 
doubts which are not dispelled when rotational Doppler shifts are related to 
the resolution of the MUSICOS spectrograph on the Pic-du-Midi used by K10. The 
rather moderate resolution of the spectra, viz. R = 35000, translates to 
$0.14$\,{\AA} FWHM for the iron lines used by K10. If one takes a spot of 
$20\degr$ diameter whose leading edge is exactly on the limb and the trailing 
edge thus $70\degr$ from the line of sight, in terms of Doppler shift the 
difference in wavelength between leading and trailing edge is a mere 
0.018\,{\AA}, corresponding to 1/8 of the resolution of the spectrograph.

\subsection{The signal of a high-contrast magnetic spot}

There can thus be hardly any doubt that the signal of a small-scale magnetic 
spot -- even when the contrast to the surroundings is high -- will be 
exceedingly faint and washed out due to the limited resolution of the 
spectrograph. The question immediately arises whether such hypothetical spots 
are observable at all - remember that they tend to appear and to disappear 
with the regularisation parameter - given that any error in the assumed 
atmospheric model translates to erroneous synthetic Stokes profiles on which 
abundance and magnetic field estimates are based.

We have for simplicity's sake assumed a centred dipole oblique rotator 
model with $90\degr$ obliquity, seen equator-on, a polar field strength of 
4.0\,kG, and a spot of $20\degr$ diameter (indeed a small-scale structure) 
at $10\degr$ latitude. Based on this geometry, we then established 2 
phase-dependent sets of Stokes spectra with an iron abundance of 
$A({\rm Fe}) = 10.00$ in the spot, and with $A({\rm Fe}) = 8.50$ over the 
rest of the stellar surface. One set uses, in the ``usual'' way, the same 
atmospheric model outside and inside the spot, with the global field strength 
distribution given by the oblique rotator model, but multiplied by 1.0, 1.5, 
and 2.0 respectively inside the spot. For the second set we adopted the
correct approach, using the appropriate atmospheric models corresponding to 
the assumed abundances, i.e. a $A({\rm Fe}) = 10.00$ Atlas12 model for the 
spot, and again calculated models with a 1.0, 1.5, and 2.0-fold enhancement 
of the magnetic field inside the spot.

In the right-hand side panels of Fig.\,\ref{fine}, we display differential 
Stokes~$I$ and $Q$ profiles at phase 0.15, differential Stokes~$V$ profiles 
at phase 0.25 (at these phases the respective signals due to the spot are 
largest), all of them with a spectral resolution of R = 35000. The differential 
signal in Stokes~$U$ is much weaker for the adopted geometry than in $Q$ or 
$V$, and therefore not plotted. The differences are given in the sense 
field enhanced spot minus normal field spot. The full lines pertain to the
``usual'' model, the dash-dotted lines to the correct model, for an
enhancement of 50\% (green, sky-blue) and 100\% (peach, red). Zeeman Doppler 
mapping addicts will find the maximum differential signal in Stokes\,$Q$ 
from a spot in which the magnetic field is increased from 2\,kG to 3\,kG 
disappointingly low, viz. a mere 0.10\% for the correct model, barely 0.07\% 
for the ``usual'' model. The situation is almost identical in Stokes\,$V$.
Even a staggering field increase in the spot from 2\,kG to 4\,kG is
reflected in no more than a 0.22\% differential signal in Stokes\,$Q$ and 
$V$, 0.15\% for the ``usual'' model. Note that these differential signals 
are not only disturbingly small but that figure~5 of K10 shows quite a few 
$Q$, $U$ and $V$ profiles which are not fitted to that accuracy!

Even at almost infinite resolution -- left side panels of Fig.\,\ref{fine} 
-- a 50\% increase in field strength in the high-abundance spot adds at 
most 0.22\% to the Stokes $Q$ signal, and 0.18\% to the $V$ signal, taking 
the correct 2 atmospheric models. Here we have plotted the Stokes profiles
relative to a reference spectrum calculated for a homogeneous iron 
abundance of $A({\rm Fe}) = 8.50$ over the entire star, in order to reveal 
the following profoundly worrisome fact: adopting the ``usual'' approach to 
the modelling of spotted stars leads to errors in all 4 Stokes parameters. 
This is particularly well visible in Stokes~$I$, resulting in spurious 
abundance estimates, but Stokes~$Q$ and $V$ too turn out to be seriously 
affected, giving magnetic field estimates in error by more than 30\%!
Looking for example just at the Stokes $V$ signal, the ``usual'' curve for 
a 4\,kG field inside the spot is almost indistinguishable from the correct
curve for a 3\,kG field. Put it the other way round, one would considerably 
overestimate both magnetic field strength and abundance in the spot using 
the ``usual'' incorrect mean model approach.

Our results clearly demonstrate that one cannot beat the limited resolution
of a spectrograph by using Stokes $Q$ and $U$ profiles in order to obtain
magnetic maps at ultra-high resolution  It also becomes obvious that any
reliable abundance and/or magnetic field estimates depend on the use of
the correct atmospheric models for the spotted stellar surface. When
these prerequisites of sufficient resolution and appropriate atmospheric
models are not met, how then can one credibly establish secondary maxima 
or exceedingly small minima, in particular in places where the magnetic 
field strength drops from about 1\,kG to a few hundred Gauss?

\section{Mapping a spotted star}
\label{spotted}

The observed Stokes~$I$ profile of a rotating star is given by the integral over the 
properly Doppler-shifted local line and continuum intensities. A spot with enhanced 
abundances normally shows up as a bump, a dip, or an asymmetry of the profile that 
varies with rotational phase. In Doppler mapping this intensity information as a 
function of wavelength is used for the reconstruction of the surface abundance 
distribution, translating each wavelength difference (relative to the line centre) 
to a velocity swath on the stellar disk. As it turns out, this inversion is highly 
vulnerable to errors in the assumed atmospheric model.

\begin{figure}
\includegraphics[width=84mm]{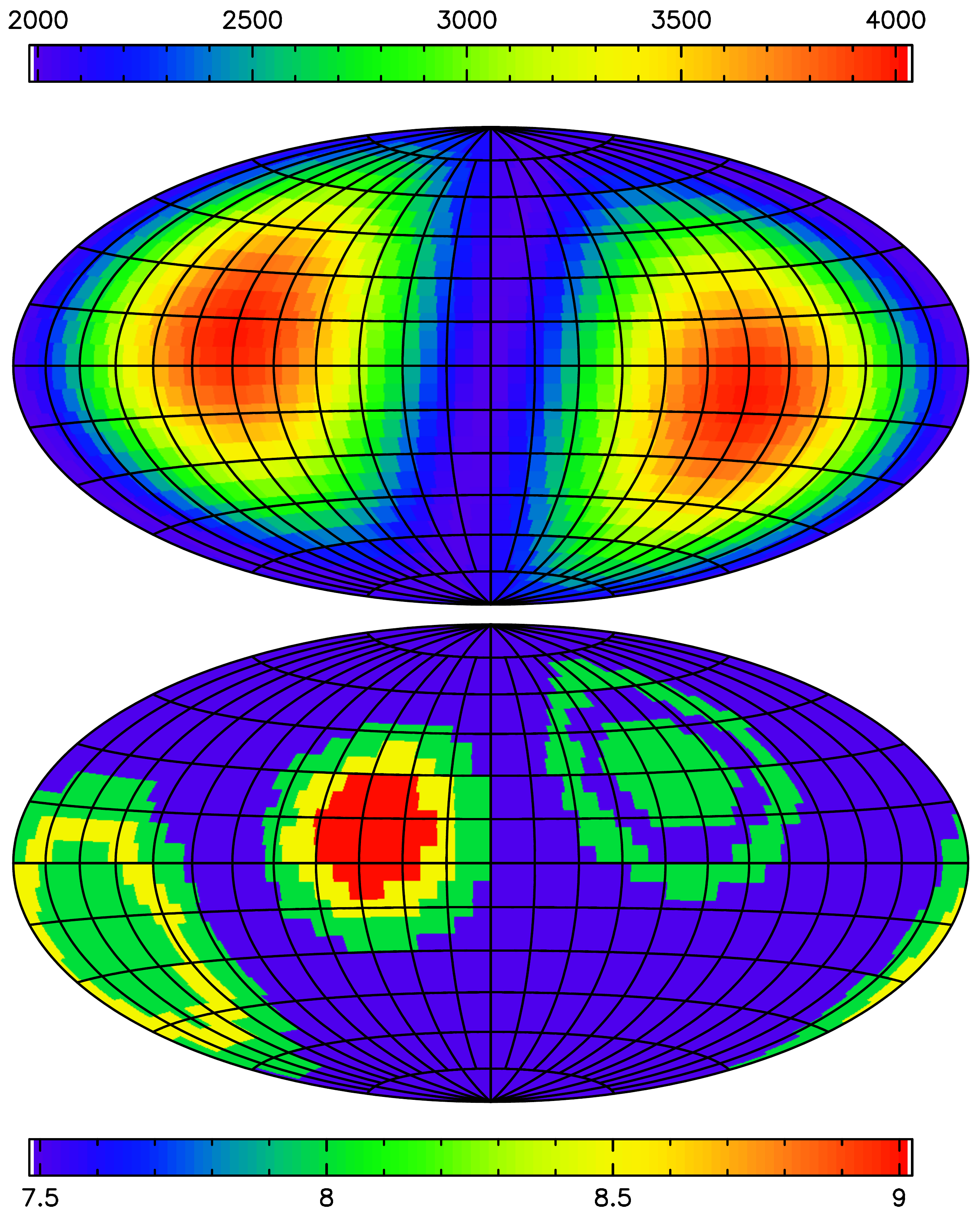}
\caption{Iron abundance pattern (lower panel) and absolute field strength 
(upper panel) adopted for the modelling of phase-dependent Stokes spectra 
of a spotted star used in subsequent ZDM discussed in section\,\ref{spotted}.}
\label{field}
\end{figure}

To illustrate this point, we consider the profile differences presented in 
Fig.\,\ref{Stokes-IV}. The largest deviation in Stokes~$I$ is found near 
4923.78\,{\AA}, about 9\,km\,s$^{-1}$ from the unshifted line position. 
This means that halfway from the disk centre to the limb, we get an 
estimate of the intensity which is incorrect by about 15\% when using 
the ``usual'' model. In Stokes~$V$ the errors can become even larger, 
up to 30\% for the dash-dotted red curve near 4923.65\,{\AA}. Of course 
such differences cannot be converted directly to abundance or field 
corrections in some region of the stellar surface because Doppler mapping 
involves global optimisation involving a regularisation function. Quite 
some time ago, Stift (1996) has pointed out that {\em ``spurious abundances 
... constitute the sometimes entirely unphysical response of a particular 
regularisation function to the spectral signature of the magnetic field''} 
but it is clear that the same holds true for errors in the atmospheric 
structure of a spotted star.

\begin{figure}
\includegraphics[width=84mm]{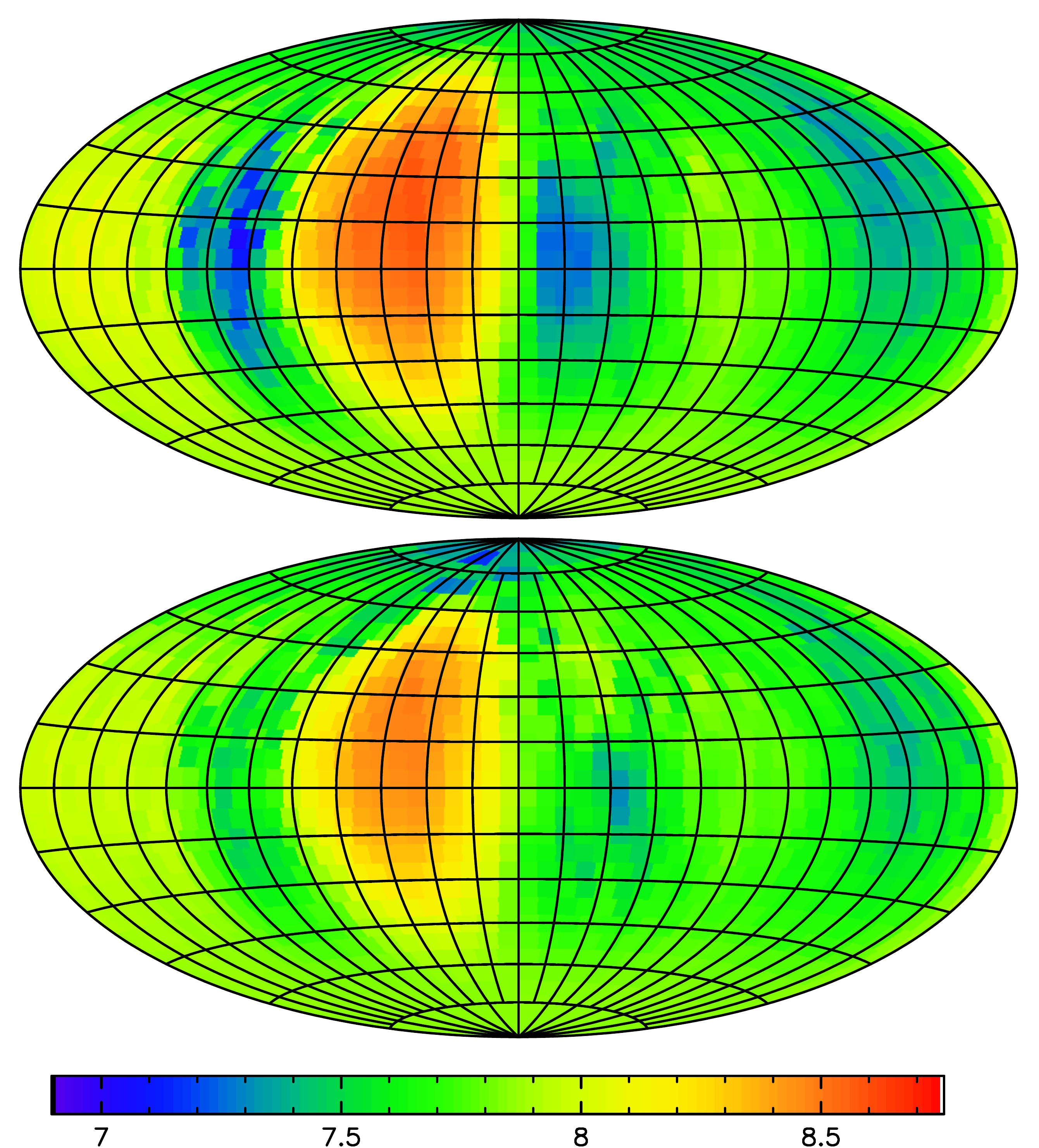}
\caption{Hammer equal-area projection of abundance distributions derived by 
Doppler mapping from 20 phase-dependent synthetic Stokes~$I$ profiles of the 
Fe\,{\sc ii} line at 4923.927\,{\AA}. Synthetic profiles taken for input have 
been calculated for a star covered by 3 spots as shown in Fig.\,\ref{field}; the 
iron abundance is up to $A({\rm Fe}) = 9.00$ in the spots and $A({\rm Fe}) = 7.50$ 
in the remaining atmosphere. One set of profiles is based on the correct 
atmospheric models, the other set on the ``usual'' approach with a mean model 
of $A({\rm Fe}) = 8.00$. The latter is employed in both inversions which lead to 
the maps shown in the respective lower (``usual'' profiles) and upper (correct 
profiles) panels.} 
\label{maps}
\end{figure}

Let us demonstrate the validity of our argument with actual Doppler maps. Using 
the revived and refined Doppler mapping code of Stift (1996), we have inverted 
a series of Stokes~$I$ profiles corresponding in resolution, wavelength coverage
and number of phases to the profiles used by K10. The star is inclined by 
$65\degr$ and covered by 3 spots as shown in Fig.\,\ref{field}; it rotates at 
$v \sin i = 18$\,km\,s$^{-1}$, and the absolute strength of the dipolar magnetic 
field ranges from 2\,kG to 4\,kG. Inside the spots the abundance can be as high 
as $A({\rm Fe}) = 9.00$, outside it is $A({\rm Fe}) = 7.50$. The ``usual'' 
intensity profiles have been calculated with a $A({\rm Fe}) = 8.00$ atmosphere, 
the correct profiles with the appropriate $A({\rm Fe}) = 7.50, 8.00, 8.50$ and 
$9.00$ atmospheres.

The inversion of a single line as in Kochukhov et al. (2004) leads to the results
shown in Fig.\,\ref{maps}. The lower panel shows an equal-area Hammer projection 
of the map obtained with the ``usual'' profiles and the ``usual'' Doppler mapping
approach, viz. under the assumption of a mean atmosphere inside and outside the 
spot, in this case a $A({\rm Fe}) = 8.00$ atmosphere. In the upper panel we have 
inverted the correct profiles in exactly the same ``usual'' way. Both inversions 
result in a comparable quality of the fit to the input profiles, both inversions 
display respective abundance maps that do not look unreasonable at all. It
is obvious that the top part of Fig.\,\ref{maps} does not resemble the bottom 
part of Fig.\,\ref{field}, but we won't dwell here on the problem of finding the 
true extensions of spots, their contrast and number (for this see e.g. Khokhlova 
1976, Goncharskii et al. 1982). We rather want to concentrate on pointing out
the mapping errors introduced by the differences between the metallicity assumed 
for the ``usual'' inversion and the actual metallicities in the spotted star. 
Note the conspicuously different respective extensions of the main spot and the 
spurious structure showing up in several places. 

\begin{figure}
\includegraphics[width=84mm]{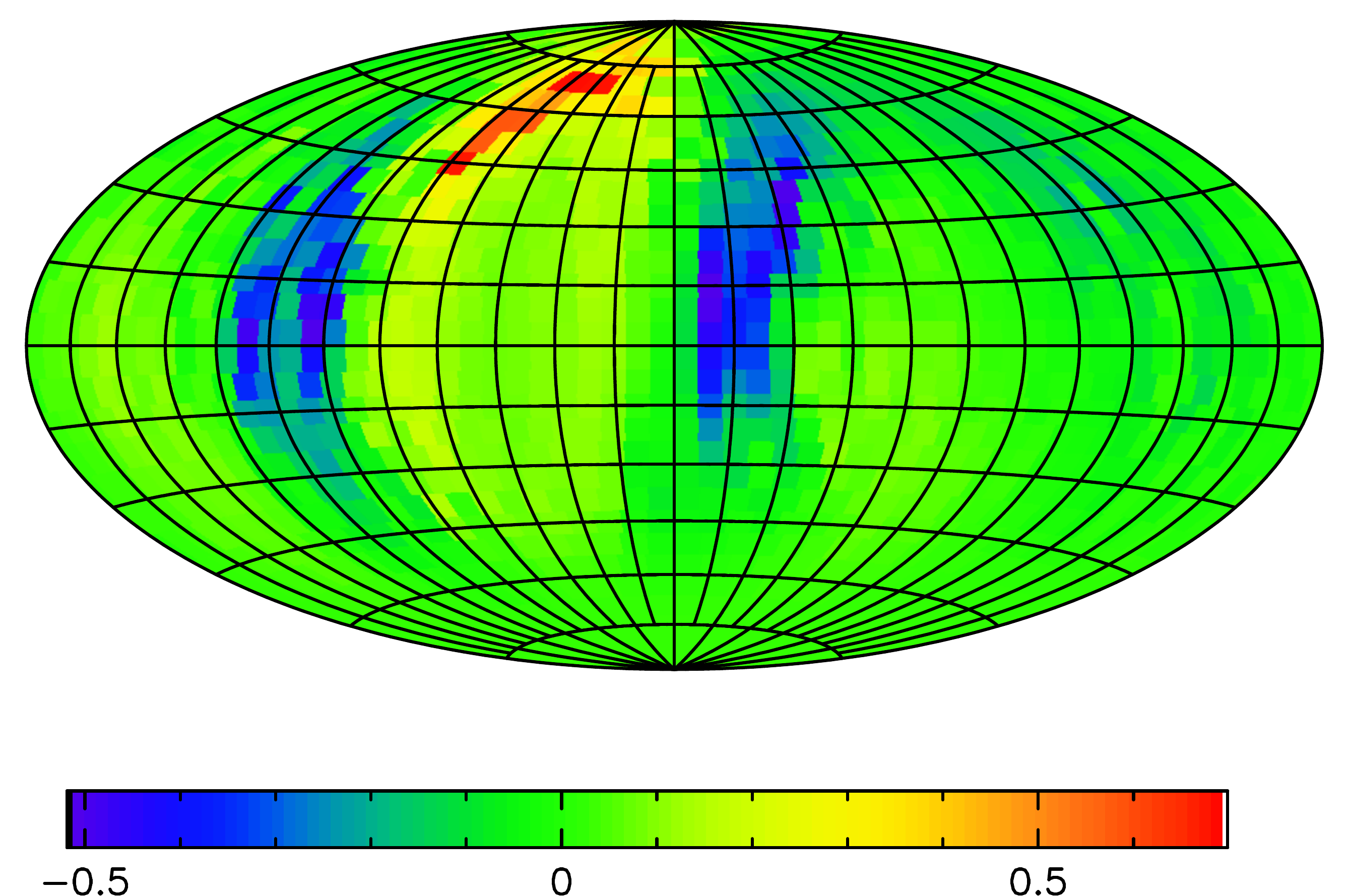}
\label{discrep}
\caption{Same data as in Fig.\,\ref{maps} but displaying the difference (dex)
between the respective abundance maps in the sense ``correct'' minus ``usual'' 
line spectrum.}
\end{figure}

To make it clear how serious the problem facing Doppler mapping really is,
we plot the difference between the map obtained with the correct profiles and
the ``usual'' map. As revealed in Fig.\,\ref{discrep}, the differences between
the respective abundances go from -0.5\,dex to almost +0.7\,dex, a remarkably 
large range given the relatively small error in the assumed abundance of the 
stellar atmosphere over most of the stellar surface. Note the spurious gradients 
near the borders of the actual strong spot.

\section{Conclusions and outlook}

It is not easy to assess which of our findings will have the most sobering
effect on (Zeeman) Doppler mapping enthusiasts keen on claiming to have found 
small-scale and high-contrast structure on the surfaces of magnetic Ap stars.
What we would certainly consider most disturbing is the fact that despite the 
use of high S/N ratio full Stokes~$IQUV$ spectra, Zeeman Doppler mapping with 
Tikhonov regularisation as carried out for example by K02 and K10 can lead to 
abundance distributions and magnetic field maps that are demonstrably and 
unequivocally incorrect. This singularly strong statement is justified 
by the following considerations: the far-reaching claims by K10 are not based
on tolerable approximations to the atmospheric structure of spotted stars 
since K10 not only neglect well-established physics of stellar atmospheres 
explored decades ago by Chandrasekhar (1935), but they also discard among 
others recent well-founded findings and recommendations made by Khan \& 
Shulyak (2007). There can be no doubt that complex Zeeman Doppler maps with 
high-contrast abundance distributions and small-scale magnetic features can 
by no means be correct when local Stokes\,$I$ profiles are in error by as 
much as 10\% of the continuum and when field estimates based on local 
Stokes\,$Q$ and $V$ profiles are in error by 30\% and more. The fate of 
the K10 paper is shared by a number of similar analyses (see e.g. 
L{\"u}ftinger et al. 2010). Having been made to believe that full Stokes 
profiles provide sufficient information not only to map 2-D abundance 
and magnetic field distributions but also 3-D abundance stratifications 
(L{\"u}ftinger et al. 2008, but see Stift \& Alecian 2009) we were rather 
surprised to see that excellent ZDM codes like I{\sc nvers}10 used by K10 
can yield good fits to the observed Stokes~$IQUV$ profiles even when the 
adopted local line profiles are grossly in error. The results shown in 
Fig.\,\ref{discrep} were equally surprising since they reveal that the 
adoption of an atmospheric model of mean metallicity in the Doppler mapping 
procedure -- the respective correct models for the star and its spots differ 
by at most $\pm 1.00$\,dex from this mean model -- leads to local abundance 
discrepancies of between $-0.50$\,dex and $+0.70$\,dex. Considering the 
fact that at $A({\rm Fe}) = 8.00$ metallicity effects are still relatively 
small and that they increase drastically for $A({\rm Fe}) > 9.50$ (see 
Fig.\,\ref{atmos} and Fig.\,\ref{blanket}), we are not overly optimistic as 
e.g. to the reliability of the iron distribution derived by K10, where the 
adopted atmosphere with $A({\rm Fe}) = 8.50$ differs from the ``local'' 
atmospheres by as much as 2\,dex.

Let us also dwell a bit on a previously unnoticed, but not altogether
surprising, complication that must be taken into account when applying ZDM.
The adoption of a mean atmosphere for a spotted star will not only lead to 
erroneous surface abundance patterns but also to incorrect magnetic 
geometries. In fact, it turns out that the Stokes $Q$, $U$ and $V$ profiles 
suffer considerably from metallicity related effects; the adoption of a mean 
atmosphere for a spotted star can induce spurious changes of about 30\% and 
more in the maximum polarisation signal. These wavelength- and phase-dependent 
profile changes will necessarily translate to errors in the distribution of 
magnetic field strength and direction. Given all the systematic errors in 
the 4 Stokes parameters which result from the use of a mean atmosphere, 
it could then prove premature to combine abundance and magnetic maps in an 
effort to verify how these data match theoretical predictions from the 
diffusion model (Michaud 1970) or to correlate the radial field field and 
abundance distributions of various elements as done by K02.

Even less surprisingly, it is shown that the existence of ``small-scale 
magnetic structures" and of ``high-contrast magnetic spots'' cannot at 
present be ascertained at any acceptable level of confidence. Even 
apparently moderate errors in the assumed atmosphere model can lead to 
non-negligible errors in the inferred magnetic field of a small, high-contrast 
spot. Any reliable detection of such a magnetic spot would only be possible 
with a much higher spectral resolution than available to K02 or K10, and it
would require a fit to the Stokes $Q$, $U$, and $V$ profiles to at least 
$0.1-0.2$\%. One should therefore consider magnetic fine structure as for 
example shown in figure\,7 of K10 to be largely an artefact of regularisation.

In view of our results, the outlook for ZDM is a bit bleak but not hopeless.
Success in the quest for more reliable maps of CP stars depends on the ability 
to eschew the use of a mean atmosphere, replacing it with a whole grid of 
stellar atmospheres with different abundances of all the elements important
enough to influence the temperature structure of the atmosphere (see 
again Khan \& Shulyak 2007). As experience with C{\sc ossam}S{\sc pot} shows 
this might put serious strain on computing resources but there is no way 
around it. In return, abundance maps based on the correct atmosphere models 
might exhibit less spectacular amplitudes and the contrast might diminish, 
reducing the complexity of the computations. It is clear that ZDM is still 
in its infancy but multi-core computer architectures in combination with 
modern object-oriented programming paradigms and powerful compilers will make 
it possible to enter the next phase. Keep however in mind that this approach 
is still full of (it is hoped) acceptable approximations and that more 
difficulties are lurking behind the bend: the influence of magnetic fields 
on the atmospheric structure, stratification, 2D and 3D stellar atmospheres 
for latitude- and longitude-dependent stratifications, etc.

\section*{Acknowledgments}

Thanks go to AdaCore for providing the GNAT GPL Edition of its Ada2005 compiler.

\bsp

{}

\label{lastpage}

\end{document}